# Soliton explosion driven multi-octave supercontinuum generation by geometry-enforced dispersion design in antiresonant hollow-core fibers

Rudrakant Sollapur[1+], Daniil Kartashov[1+], Michael Zürch[1,2+], Andreas Hoffmann[1], Teodora Grigorova[1], Gregor Sauer[1], Alexander Hartung[3], Anka Schwuchow[3], Joerg Bierlich[3], Jens Kobelke[3], Markus A. Schmidt[3,4], Christian Spielmann[1,5*]

[1]*Institute of Optics and Quantum Electronics, Abbe Center of Photonics, Friedrich Schiller University, Max-Wien-Platz 1, 07743 Jena, Thuringia, Germany*

[2]*University of California, Chemistry Department, D39, Hildebrand Hall, CA 94720 Berkeley, California, USA*

[3]*Leibniz Institute of Photonic Technology e.V., Albert-Einstein-Str.9, 07745 Jena, Thuringia, Germany*

[4]*Otto Schott Institute of Material Research, Abbe Center of Photonics, Fraunhoferstr.6, Friedrich Schiller University, 07743 Jena, Thuringia, Germany*

[5]*Helmholtz Institute Jena, Helmholtzweg 4, 07743 Jena, Thuringia, Germany*

[+] These authors contributed equally to the work
[*] e-mail: christian.spielmann@uni-jena.de

## ABSTRACT

Ultrafast supercontinuum generation in gas-filled waveguides is one enabling technology for many intriguing application ranging from attosecond metrology towards biophotonics, with the amount of spectral broadening crucially depending on the pulse dispersion of the propagating mode. Here we show that the structural resonances in gas-filled anti-resonant hollow core optical fiber provide an additional degree of freedom in dispersion engineering, allowing for the generation of more than three octaves of broadband light ranging deep UV wavelength towards the near infrared. Our observation relies on the introduction of a geometric-induced resonance in the spectral vicinity of the pump laser outperforming the gas dispersion, thus yielding a dispersion being independent of core size, which is highly relevant for scaling input powers. Using a Krypton filled fiber we observe spectral broadening from 200 nm towards 1.7 µm at an output energy of about 23 µJ within a single mode across the entire spectral bandwidth. Simulations show that the efficient frequency generation results from a new physical effect – the soliton explosion – originating from the strongly non-adiabatic mode dispersion profile. This effect alongside with the dispersion tuning capability of the fiber will enable compact ultrabroadband high energy sources spanning from the UV to the mid-infrared spectral range.

## INTRODUCTION

Since the very first observation of supercontinuum generation in bulk crystals or glasses in the 1970s [1, 2], this phenomenon has been the subject of numerous investigations and represents one of the core physical problems in ultrafast nonlinear optics. Supercontinuum generation in optical fibers was first demonstrated in [3], but the advent of photonic crystal fibers (PCFs) in 1999-2000 [4-8] revolutionized the field. PCFs support highly nonlinear interaction under the condition of single-mode guidance in a broad spectral range, having enabled unique light sources of temporally and spatially coherent white light which can in fact be considered as an arbitrary wavelength and bandwidth source of coherent radiation. Clearly, the importance of these sources is apparent in the wide array of applications in spectroscopy, microscopy (including beyond-diffraction-limit methods like Stimulated Emission Depletion Microscopy (STED) [9]), attosecond metrology [10], frequency comb technology [11-14], telecommunications, biophotonics [15] as well as semiconductor inspection [16]).

A significant advantage in terms of achievable energy of the spectral supercontinuum and flexibility in the control over the net dispersion characteristics of PCF was achieved with the invention of the hollow-core PCF and especially Kagome fiber [17, 18]. A Kagome fiber is a hollow-core PCF combining the advantages of conventional hollow-core fiber (i.e., capillary [19]) with low transmission losses in much broader spectral bandwidth than in solid-core PCFs. Using gas as nonlinear medium inside the hollow core, the energy of the generated supercontinuum has been increased by orders of magnitude due to the higher optical breakdown threshold of gases in comparison to solids and the ability to fine tune the dispersion by changing the gas pressure. Also, new regimes of generation at the intensities near the breakdown threshold and involving highly nonlinear process of gas ionization, enhancing the short wavelength spectral region of the supercontinuum, become accessible [20, 21].

The physics of supercontinuum generation in PCF is very rich and complex and a subject of intensive research up to date. A detailed review on experimental and theoretical results can be found, for instance, in [22] and [23]. The whole complexity of supercontinuum generation by femtosecond laser pulses in PCF can be divided into two distinct cases determined by the sign of the group velocity dispersion (GVD), which correlates with the quadratic term in a Taylor expansion of the modal propagation constant. In the normal dispersion regime (positive sign of GVD) the spectral broadening is governed primarily by self-phase modulation (SPM) and self-steepening [24]. In the case of anomalous dispersion (negative sign of GVD), supercontinuum generation results from solitonic dynamics, in particular solitonic pulse self-compression, soliton fission and dispersive wave emission [22, 25, 26].

In this contribution we report on very efficient supercontinuum generation in a new type of hollow-core PCF – the so-called anti-resonant hollow core fibers (ARHCF) – providing an additional degree of freedom in dispersion management and opening up a new regime of solitonic dynamics which we refer as *soliton explosion*. Pumping such an ARHCF filled with Krypton gas in the spectral vicinity of the fundamental strand resonance (resonance imposed by the glass ring surrounding the core section) with femtosecond pulses, an ultrabroadband (more than three octaves) supercontinuum between the UV and near-infrared (NIR) with more than 10 µJ pulse output energy is generated. Propagation in the fundamental mode is observed for all wavelengths. One key feature of the observed spectral broadening is its sudden onset with a slight increase in input pulse energy above a certain threshold value. Our numerical simulations suggest that the mechanism behind the supercontinuum generation is a novel type of non-adiabatic soliton dynamic, caused by the strong resonance of the submicron glass strand surrounding the hollow core, imposing the group velocity dispersion to change by orders of magnitude several times from anomalous to normal dispersion with multiple zero dispersion wavelengths within a narrow spectral interval.

**MATERIALS AND METHODS**

**Anti-resonant hollow core fibers**

The fabrication of the ARHCF used in the experiments relies on stacking six thin-walled capillaries (outer diameter 3.7 mm, wall thickness 0.4 mm) into a jacket tube and drawing this arrangement into cane and fiber. Details of the fabrication procedure can be found elsewhere [27].

The overall working principle of anti-resonant hollow core fibers relates back to the fact that dielectric interfaces can show reflection values close to unity in case of grazing incidence [28]. Even higher reflectivity values are obtained for two parallel interfaces (i.e., thin films), as interference effects strongly modify the interface reflection [27]. In case the film thickness is of the order of the exciting wavelength, a resonance can cause light to tunnel through the glass strand. The spectral positions of these resonances are mainly defined by the strand thickness *t:*

$$\lambda_m = 2t\sqrt{n^2 - 1}/m \qquad (1)$$

Here the incident medium is air, *n* the refractive index of glass and *m* the mode order [29]. Between two adjacent resonances, very high reflectivity values are obtained, which is employed in anti-resonant reflecting optical waveguides (ARROW), e.g., microfluidics [30] and more recently within hollow core fibers [31, 32]. The supported leaky modes are effectively guided inside a thin ring of glass, whereby the modal attenuation scales with the inverse of the cubed air core radius [28] similar to a capillary [33] but with dramatically reduced losses (a few dB/m for core sizes of about 50 µm instead of tens of dB/m for a hollow-core fiber of the same diameter).

As the group velocity dispersion depends on the second derivative of the propagation constant with respect to frequency, it is possible to massively vary the pulse dispersion properties in the vicinity of a strand resonance compared to a simple capillary or Kagome type fibers (Fig. 1b and inset of Fig. 1b). By adjusting strand thickness (see equation. (1)) and core size (equations (2 and 3)) it is in fact possible to generate four zero-dispersion wavelengths within any desired spectral range in the transmission window of the fiber, offering great potential for dispersion engineering and power scaling beyond any of the existing schemes. Here we show that dispersion engineering via choosing the correct position of the strand resonances offers a unique possibility for spectral tuning of supercontinuum generation, especially at IR wavelengths.

**Experiments**

The experiments have been carried out using a femtosecond Ti:sapphire laser system, delivering 80 fs pulses with energies up to 1 mJ at 0.8 µm central wavelength and 1 kHz repetition rate (Spitfire, SpectraPhysics). The detail experimental setup is shown in Fig. S1 in the Supplementary Material. The linearly polarized pulses emitted by the laser pass through a combination of thin-film polarizer and half-wave plate to control the input pulse energy from a few nJ to 50 µJ. The beam is collimated to a diameter of 12 mm and coupled into the 50 µm core, 250 mm long ARHCF (inset of Fig. 1) using a lens (focal length: 200 mm, focal beam diameter: 17 µm). A transmission efficiency of ≈65% was determined for the evacuated fiber.

This value remained constant in the whole range of gas pressures and laser energies used in the experiments. The fiber input and output are mounted inside independent gas cells allowing either for a constant gas pressure along the fiber or a pressure gradient. The output emission, attenuated by reflection from a wedge, is focused onto the entrance slits of three different spectrometers: two silicon CCD spectrometers (Ocean Optics USB4000-UV-VIS and USB2000+) for characterization in the UV and VIS spectral range and an InGaAs CCD spectrometer (Ocean Optics NIRQuest512) for the NIR spectral range. This combination of spectrometers allows spectral characterization between 200 nm and 1.7 µm, whereas the full spectrum is obtained by stitching together the individual spectra within the overlap regions (800 and 1000 nm). Each individual spectrum is corrected for the spectral distribution of the respective spectrometer sensitivity. The output beam profiles at different wavelengths are recorded using a suitable combination of narrowband interference filters and cameras (UV/VIS: Coherent LaserCam-HR; NIR: InGaAs ABS GmbH, IK-1513) placed at a distance of ≈120 mm from the fiber end.

**Numerical simulations**

In simulations we solve numerically the Unidirection Propagation Pulse Equation (UPPE) in frequency domain [34]:

$$\frac{\partial \mathcal{E}}{\partial z} = i(k(\omega,p) - \omega/v)\mathcal{E} + i\frac{\omega^2}{2c^2 k(\omega,p)}\chi_3 \mathcal{F}[E^3]$$

Here $\mathcal{E}(\omega, z) = \mathcal{F}[E(t,z)]$ is the spectral amplitude, $\mathcal{F}$ is the Fourier transform operator, $E(t,z)$ is the real electric field in the laser pulse, $k(\omega, p)$ is the wavevector determined via the dispersion relation (3) where the pressure dependence of gas index of refraction is described as $n_g(\omega,p) = \left(2\frac{p}{p_0}\frac{n_0^2-1}{n_0^2+2} + 1\right)^{1/2} \left(1 - \frac{p}{p_0}\frac{n_0^2-1}{n_0^2+2}\right)^{-1/2}$, with $p_0$ being the normal atmospheric pressure $n_0(\omega)$ is refractive index of the gas at atmospheric pressure calculated from Sellmeier equation for Kr [35], $v$ is the velocity of a reference frame chosen for convenience of simulations, and $\chi_3$ is the third order susceptibility which is assumed to be instantaneous and, hence, frequency independent.

**RESULTS AND DISCUSSION**

**Experimental results**

The conceptual scheme of the experiment is shown in Fig.1a. The fiber used in the experiments consists of a central hollow core (diameter: 50 µm, inset of Fig. 1a) surrounded by a thin silica ring being supported by a series of connecting strands (total fiber outer diameter 125 µm). The small thickness of the strands (average thickness 495 nm) leads to a fundamental resonance (*m* = 1) at around 1.02 µm (Fig. 1b), which is on the long wavelength-side of the central wavelength of the used ultrafast laser (0.8 µm). Cut-back measurements reveal a series of transmission bands with losses between 5 and 10 dB/m across the UV, VIS and NIR spectral domains. Insufficient spectral density of the used light source (halogen lamp) prevents us from quantifying the loss below 350 nm, whereas the raw data indicates two more transmission bands with a minimum transmission wavelength of 220 nm. It is important to note that the used fiber reveals a variation of the strand thickness of about 10% resulting from various fabrication issues, which leads to a broad bandgap (Fig. 1c) instead of a narrow band resonance (Fig. 1b).

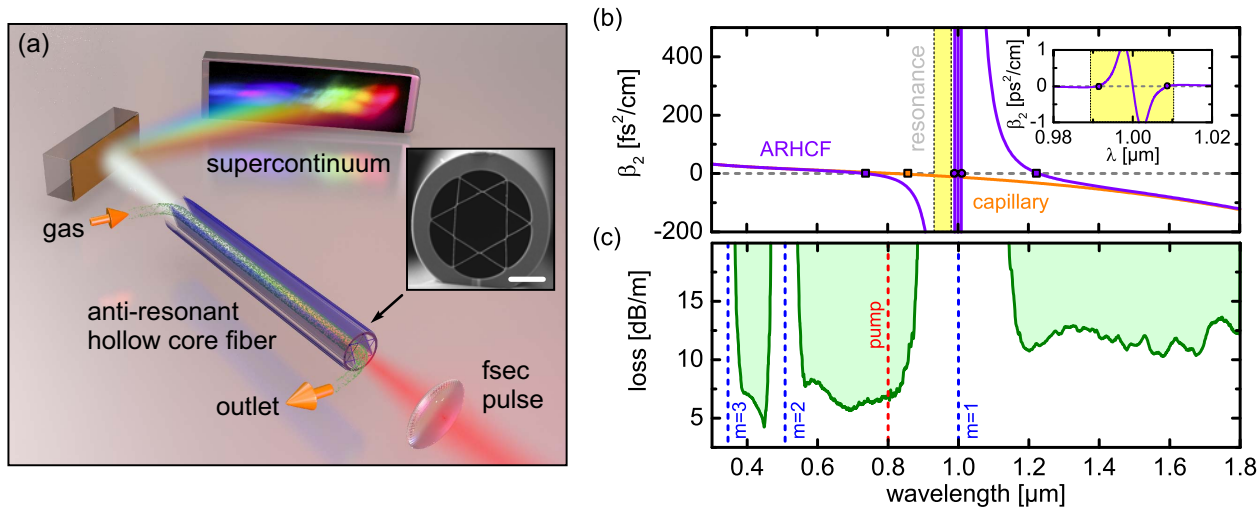

**Fig. 1:** a) Conceptual scheme of the experiment. The inset image shows the anti-resonant hollow core fiber used in the experiments (The white scale bar indicates a length of 20 µm). (b) Simulation of the frequency dependence of the group velocity dispersion of the fundamental mode in the ARHCF filled with Krypton at a pressure of 6 bar (purple curve, strand resonance located at 1 µm; orange: corresponding gas-filled capillary dispersion when the strand resonance is neglected). The squares mark the zero group-velocity dispersion wavelengths. The inset is a close-up view of the GVD in the vicinity of the strand resonance. (c) Measured spectral distribution of the loss of the fundamental core mode of the anti-resonant hollow core fiber shown in (a). The vertical blue dashed lines indicate the spectral positions of the resonances according to formula (1) and the vertical red dashed line refers to the central wavelength of the femtosecond pump laser.

The dependence of spectral broadening of 80 fs laser pulses at 0.8 µm wavelength on input pulse energy in the case of the ARHCF homogenously filled by Krypton at 6 bar is shown in Fig. 2b-f. Clearly visible is a sudden onset of the spectral broadening at around 10 µJ input energy, corresponding to 6.5 µJ output energy (Fig. 2d) which evolves into a multi-octave spanning supercontinuum extending from about 200 nm to 1.4 µm at the output pulse energy of 12.1 µJ. It is remarkable that a few-hundred nanometer spectral broadening is observed with only 25% of increase in the input energy (Fig. 2d-e). At 10 µJ of the input energy, broadening from 200 nm to 900 nm is observed, whereas the signal in the NIR is below the noise limit of the near-infrared spectrometer. Another 20% increase in the input energy triggers dramatic spectral broadening towards the IR spectral range (Fig. 2c-d). A further increase in the input pulse energy leads mainly to increase in the supercontinuum spectral intensity and bandwidth.

With a constant pressure in the fiber, the maximum achieved output energy in the 3-octave supercontinuum is 12.1 µJ with ≈19 µJ input energy. Higher input energies lead to significant ionization of the gas at the fiber input and damaging of the fiber structure. At the maximum input energy used in the experiments with the constant gas pressure, measurements of the far-field beam profiles confirm fundamental mode propagation at all wavelengths of the supercontinuum (Fig. 2n-p).

To overcome gas ionization at the fiber input and increase the energy of the supercontinuum further, a differential gas pumping setup was used, keeping the input side of the fiber under vacuum (10s of mbar) and maintaining high gas pressure (up to 6 bar) at the output end. This gradient gas pressure allows avoiding of dense gas plasma at the fiber input, overall enabling higher energies to be coupled in. Additionally, tight focusing of the input beam allows avoiding direct damaging of the fiber input by the focused laser beam and enabling the long term operation with no degradation in the output beam quality. This configuration yields similar spectral broadening at the short wavelength side of the spectrum for roughly double the coupled energy, but with an extended spectral broadening towards the NIR with the largest wavelength being 1.7 µm at the output pulse energy of 23 µJ (Fig. 2a). It is worth mentioning that the high modal loss associated with the strand resonances (Fig. 1b) is effectively suppressed for the highest input energies (yellow bars in Figs. 2a,b), indicating that the nonlinear process of frequency conversion is able to efficiently fill up the high-loss spectral regions [36]. At maximum spectral broadening we estimate nearly 75% energy in the IR (>700 nm), ~24% in the VIS (400-700 nm) and ~1% in the UV (< 400 nm).

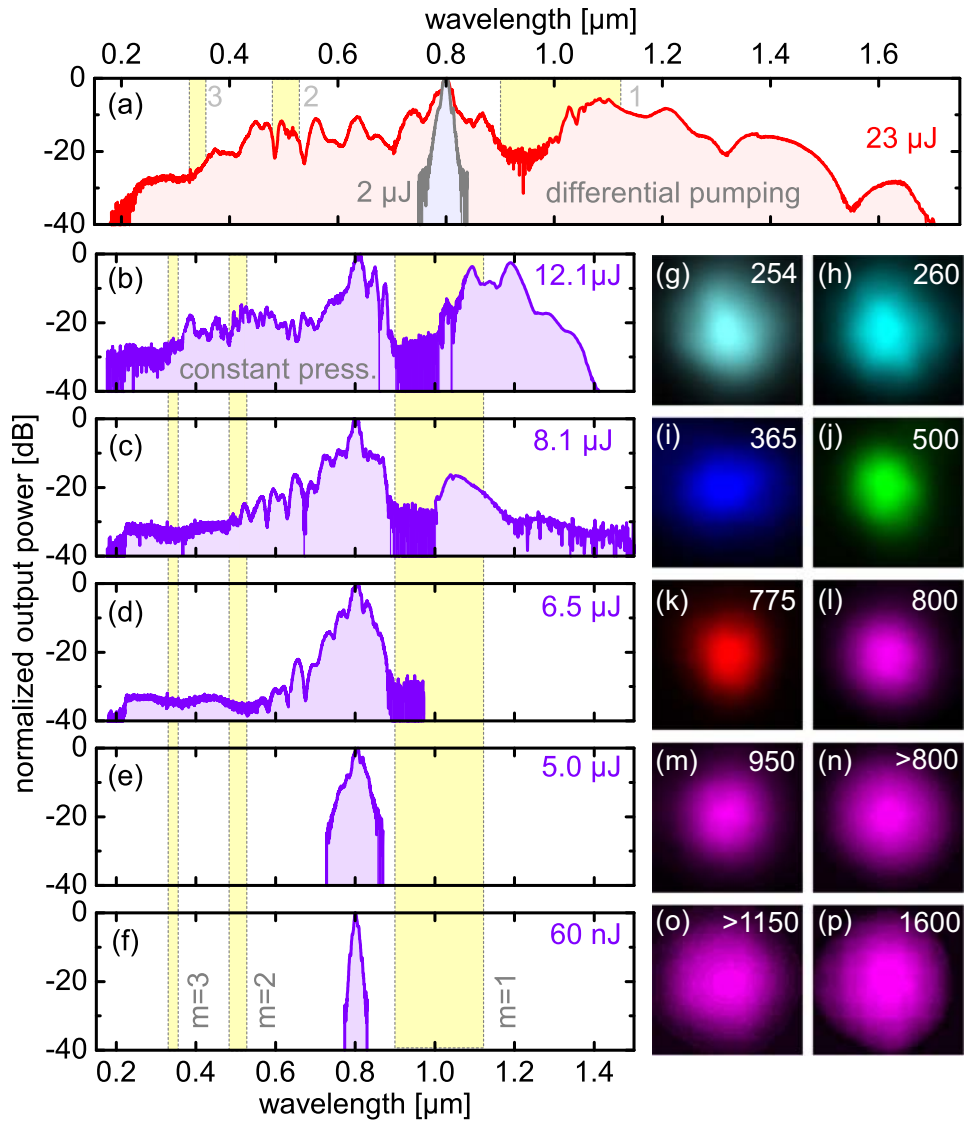

Fig. 2: Measured spectral distribution of the output light after the Krypton-filled hollow core fiber for various input pulse energies (the corresponding output energy values are given in the plots). (a) Fiber including pressure gradient (low-pressure side facing input, pressure at output: 6 bar). (b)-(f) Homogenously filled fiber at a pressure of 6 bar. (g)-(p) Far field images (in false colors) of the output mode at various wavelength for the input energy of 12.1 µJ are shown on the right-side of the figure (numbers indicate the central wavelength in nm of the band pass filters).

**Numerical Analysis**

To reveal the physical origin of the observed broadening we performed nonlinear pulse propagation simulation using the UPPE [34], assuming that the light of the pump and at all generated wavelength is in a single mode along the entire fiber length. To calculate the modes and the modal effective index of refraction, we also performed finite elements simulations for the ARHCF structure used in the experiments [32]. Far away from the strand resonance the modal dispersion of the fundamental mode can be well fitted by the dispersion relation of a gas-filled HCF [33], which was used for simulating the nonlinear pulse propagation in Kagome fibers [21]:

$$k(\omega) = \frac{\omega}{c}\left(n_g(\omega,p) - 2\left(\frac{u_{11}^2 c}{\omega d}\right)^2\right) \qquad (2)$$

with the angular frequency ω, the gas pressure *p*, the pressure dependent refractive index of the gas $n_g$ (at constant temperature), the first root of the zero-order Bessel function $u_{11} \approx 2.405$, the core diameter *d* and the vacuum speed of light *c*.

To simulate the influence of the strand resonance on supercontinuum generation process, we modified the dispersion relation (2) as:

$$k(\omega) = \frac{\omega}{c}\left(n_g(\omega,p) - 2\left(\frac{u_{11}^2 c}{\omega d}\right)^2 - \frac{a(\omega-\omega_{res})}{(\omega-\omega_{res})^2+b}\right) \qquad (3)$$

where the coefficients *a* and *b* are calculated from fitting the effective mode index in the vicinity of the resonance by the Lorentz-derivative function. The resulting spectral distributions of the GVD for a Kagome-fiber (orange curve) and for our ARHCF (purple curve) are shown in Fig. 1b. The simulations are carried out assuming 80 fs, 12 µJ laser pulses at 0.8 µm wavelength propagating in the 25 cm long, 50 µm diameter ARCHF filled by Kr under 6 bar constant pressure, similar to experimental conditions. For simplicity we have neglected the propagation losses. The strand fundamental resonance wavelength is chosen to be located at 1 µm corresponding to the middle of the measured NIR resonance band (Fig. 1c).

**Discussion**

The direct comparison of the simulations of the nonlinear pulse propagation in the Krypton-filled ARHCF with a Kagome-fiber of equal core diameter clearly shows the profound influence of the strand resonance on the bandwidth of the generated light. In the Kagome case, monotonous spectral broadening driven by SPM and self-steepening is observed, yielding a maximum bandwidth of approx. 900 nm for a 20 cm long fiber (Fig. 3b).

In contrast, the ARHCF shows a three-octave spanning broadening extending from 200 nm to more than 2 µm (Fig. 3a) with a sudden dramatic increase in the broadening rate. SPM initially drives the spectral broadening in the same way as in Fig. 3b, until the red part of the generated spectrum reaches the strand resonance after approximately 6 cm of propagation. At this point, explosion-like dynamics of spectral broadening is observed. Within 1 cm of further propagation length the spectrum expands down to 300 nm on the short and above 1.5 µm on the long wavelength side of the spectrum. Our qualitative explanation of this phenomenon is as follows: according to Fig. 1b, the GVD increases by orders of magnitude in the vicinity of the resonance, which leads to two crucial consequences. First, the increasing GVD leads to boost in solitonic dynamics and shortening of the propagation distance at which solitonic self-compression and subsequent fission will occur, since the rate of soliton fission as well as the total amount of solitons are proportional to $\sqrt{L_D} \sim 1/\sqrt{|\beta_2|}$ (L$_D$: dispersion length; $\beta_2 = \partial^2 k/\partial \omega^2$: GVD) [22]. Second, an abruptly growing GVD implies large values of higher-order dispersion terms which cause self-consistent large frequency shift of forming solitons and massive dispersive wave emission. The efficiency of dispersive wave emission is additionally increased by the presence of several spectral regions with opposite sign of GVD and multiple zero-GVD wavelengths provided by the strand resonance (Fig. 3a). This enables generation of dispersive waves towards both short- and long-wavelengths [see 22 and references within]. Therefore we propose that the observed sudden increase in spectral broadening in ARHCF is the result of strongly non-adiabatic spectral dynamics of solitons in the vicinity of a resonance accompanied by massive dispersive wave emission towards both the red and the blue sides of the spectrum. This assumption is qualitatively confirmed by the pulse dynamics in time domain presented in Fig. S2 in the supplementary material.

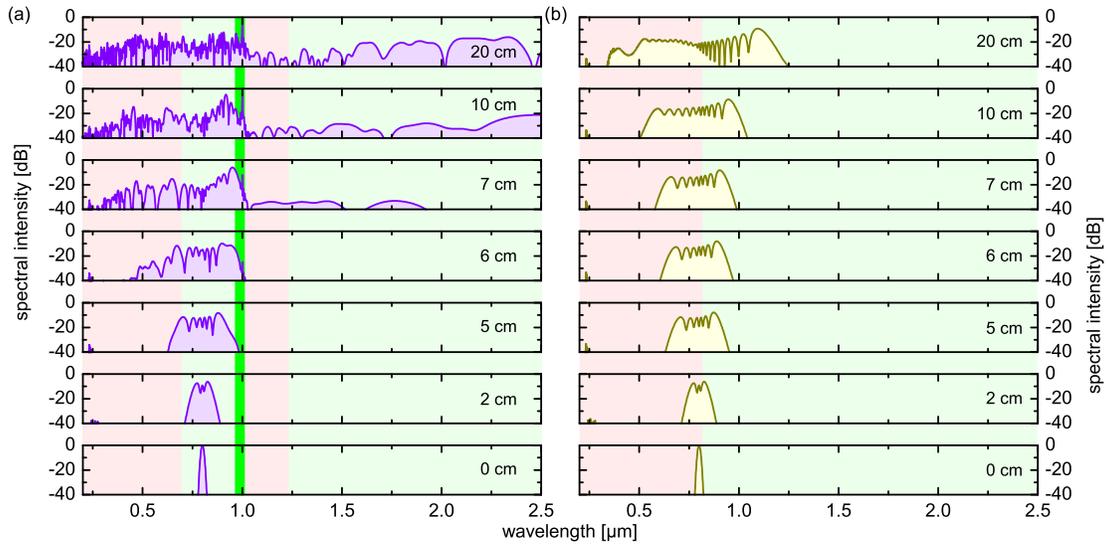

**Fig. 3: Simulation results comparing ARCHF to Kagome-type fibers.** (a) Simulation of the spectral distribution of the spectral intensity at various positions inside the ARCHF (core diameter 50 µm). The numbers on the right side of each plot indicate the propagation distance relative to the input of the fiber. The green bar refers to the strand resonance at 1 µm (Fig. 1b). The corresponding distribution for a Kagome-type HCF of the same core diameter is shown in (b). In both diagrams, the vertical lines show the position of the zero GVD wavelengths. The light magenta and green filling indicate positive and negative signs of GVD, respectively.

The numerical simulations qualitatively reproduce the experimental observations. Albeit in simulations we show the development of spectrum as a function of propagation distance, whereas in experiment the dependence on input pulse energy is presented, these two dependences are roughly equivalent: one can fix the input pulse energy and measure the spectrum for different fiber length (which is technically more challenging), or fix the fiber length and change the pulse intensity. Our numerical simulations confirm qualitative equivalence of spectral dynamics in both cases (not shown here), whereby the underlying physics is more evident in plots of Fig. 3.

**CONCLUSION AND OUTLOOK**

In this contribution we introduce the concept of modal dispersion engineering in noble gas filled anti-resonant hollow core fibers via structural strand resonances, offering a new degree of freedom to design sophisticated dispersion landscapes, allowing here to generate multiple octave spanning supercontinua from deep UV wavelengths towards the near infrared with high spectral density. By launching femtosecond pulses into a Krypton filled fiber with submicron silica strands of predefined thickness adjacent to the gas-filled core, we observe three-octave-spanning output spectra at pulse energies of 23 µJ, originating from multiple soliton fission and dispersive wave generation processes, which we refer as soliton explosion. The key to the observed broadening is the strongly varying group velocity dispersion in the spectral vicinity around the strand resonance, which outperforms the dispersion of the gas and allows to access new physics. The abruptly changing dispersion leads to a dramatic speedup in solitonic dynamics and enables a broad spectrum via multiple emitted dispersive waves. The experiments show a maximal broadening from UV wavelength (200 nm) to about 1.7 µm using differential gas pumping with fundamental-mode-type output mode at all generated wavelengths. Qualitative agreement between experiments and nonlinear pulse propagation simulations has been achieved, showing almost similar spectral behavior particular at short wavelengths. One striking feature of the generated spectra is the filling up of the high-attenuation domains of the anti-resonant fiber, diminishing the effect of strand-induced loss and leading to overall smooth spectra.

Future research will focus on a deeper understanding of the soliton explosion effect using strand resonance coupling, with the final aim to reach either deep-UV or mid-IR wavelength, whereas the latter was already indicated by our simulation. Due to its unique spectral broadness, we expect our concept to be highly relevant in application-driven areas such as spectroscopy, microscopy, metrology and biophotonics.

**ACKNOWLEDGMENT**

Funding from the federal state of Thuringia (Forschergruppe Fasersensorik; FKZ: 2012 FGR0013) and ESF is highly acknowledged. M. Z. acknowledges support from the Humboldt Foundation. R. S. acknowledges support from German Research Foundation (DFG) for funding through International Research Training Group (IRTG) 2101.

**Competing financial interests**

The authors declare no competing financial interests

*Supplementary Material for*

**Soliton explosion driven multi-octave supercontinuum generation by geometry-enforced dispersion design in antiresonant hollow-core fibers**

Rudrakant Sollapur[1+], Daniil Kartashov[1+], Michael Zürch[1,2+], Andreas Hoffmann[1], Teodora Grigorova[1], Gregor Sauer[1], Alexander Hartung[3], Anka. Schwuchow[3], Joerg Bierlich[3], Jens Kobelke[3], Markus A. Schmidt[3,4], Christian Spielmann[1,5*]

[1]*Institute of Optics and Quantum Electronics, Abbe Center of Photonics, Friedrich Schiller University, Max-Wien-Platz 1, 07743 Jena, Thuringia, Germany*

[2]*University of California, Chemistry Department, D39, Hildebrand Hall, CA 94720 Berkeley, California, USA*

[3]*Leibniz Institute of Photonic Technology e.V., Albert-Einstein-Str.9, 07745 Jena, Thuringia, Germany*

[4]*Otto Schott Institute of Material Research, Abbe Center of Photonics, Fraunhoferstr.6, Friedrich Schiller University, 07743 Jena, Thuringia, Germany*

[5]*Helmholtz Institute Jena, Helmholtzweg 4, 07743 Jena, Thuringia, Germany*

**S1 Detailed Experimental Setup**

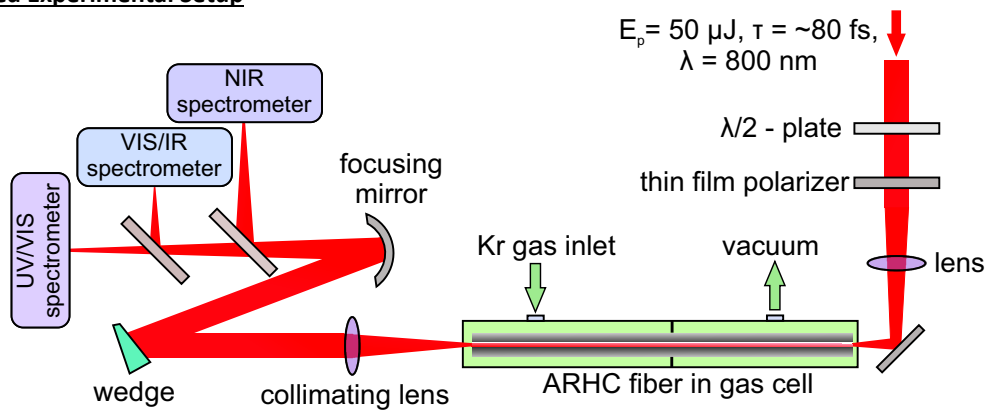

**Fig. S1:** Sketch of the experimental setup used to generate the supercontinuum. The green arrows refer to the situation of differential pumping for creating a pressure gradient inside the fiber sample.

---

[+] These authors contributed equally to the work
[*] e-mail: christian.spielmann@uni-jena.de

## S2 Spectrum and Time domain simulation for ARHCF

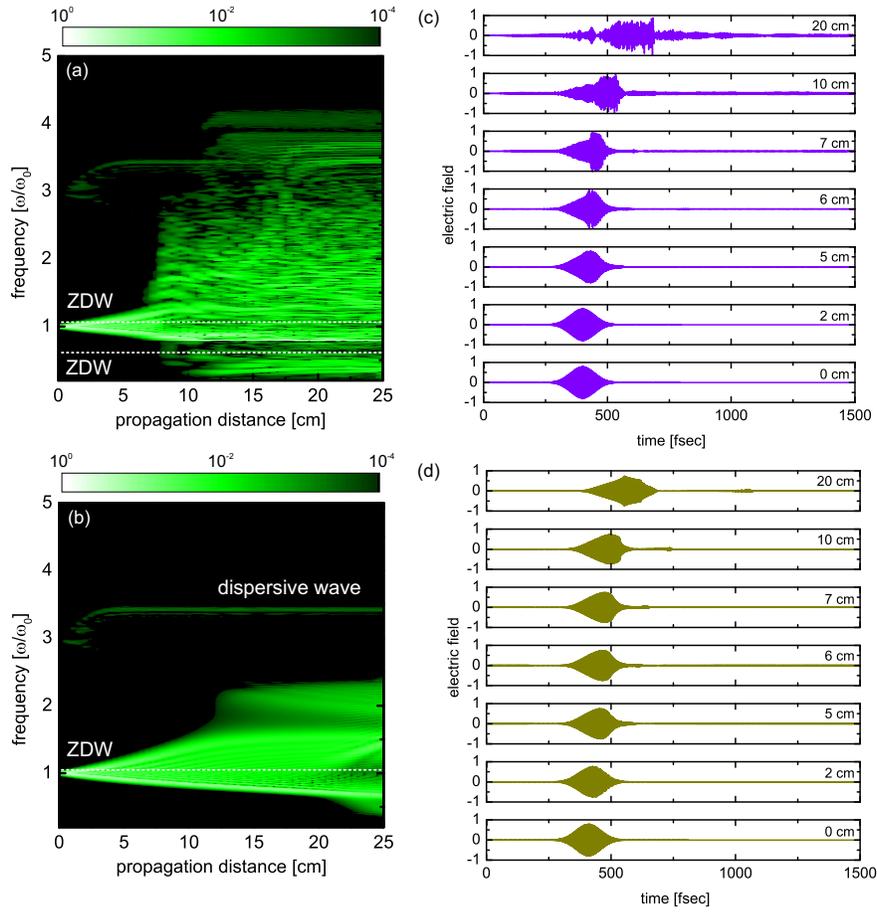

**Fig. S2:** Spectral and time domain picture for ARHCF and Kagome fiber. Spectral evolution of laser pulses (see parameters in the main text) along the propagation coordinate in the ARHCF a) without a strand resonance and b) with the strand resonance located at 1 µm (marked by the red solid line). Here the frequency axis is normalized to the laser angular frequency $\omega_0$. The spectra in Fig. 3 of the main text are line profiles of a) and b) at the fiber length 20 cm. The black dotted lines mark position of zero GVD wavelengths. Panels c) and d) – temporal evolution of the laser pulse corresponding to the spectral evolution a) and b) correspondingly.

[11] Diddams SA, Jones DJ, Ye J, Cundiff ST, Hall JL. Direct link between microwave and optical frequencies with a 300 THz femtosecond laser comb. Phys. Rev. Lett. 2000; **84:** 5102-5105.

[12] Jones DJ, Diddams SA, Ranka JK, Stentz AJ, Windeler RS *et al*. Carrier-envelope phase control of femtosecond mode-locked lasers and direct optical frequency synthesis. Science 2000; **288:** 635-639.

[13] Holzwarth R, Hänsch T, Knight JC, Wadsworth WJ, Russell P. Optical frequency synthesizer for precision spectroscopy. Phys. Rev. Lett. 2000; **85:** 2264-2267.

[14] Newbury NR. Searching for applications with a fine-tooth comb. Nature Phot. 2011; **5:** 186-188.

[15] Tu HH, Boppart SA. Coherent fiber supercontinuum for biophotonics. Laser Photonics Rev 2013; **7:** 628-645.

[16] Byeon CC, Oh MK, Kang H, Ko DK, Lee J *et al*. Coherent absorption Spectroscopy with supercontinuum for semiconductor quantum well structure. J Opt Soc Korea 2007; **11:** 138-141.

[17] Couny F, Benabid F, Light PS. Large-pitch kagome-structured hollow-core photonic crystal fiber. Opt. Lett. 2006; **31:** 3574-3576.

[18] Wang YY, Wheeler NV, Couny F, Roberts PJ, Benabid F. Low loss broadband transmission in hypocycloid-core Kagome hollow-core photonic crystal fiber. Opt. Lett. 2011; **36:** 669-671.

[19] Nisoli M, De Silvestri S, Svelto O, Szipöcs R, Ferencz K *et al*. Compression of high-energy laser pulses below 5 fs. Opt. Lett. 1997; **22:** 522-524.

[20] Mak KF, Travers JC, Hölzer P, Joly NY, Russell P. Tunable vacuum-UV to visible ultrafast pulse source based on gas-filled Kagome-PCF. Opt. Express 2013, **21:** 10942-10953.
20